\documentclass[iop]{emulateapj}

\pdfoutput=1

\usepackage{graphicx}
\usepackage{color}
\usepackage{subfigure}
\DeclareRobustCommand{\ion}[2]{%
\relax\ifmmode
\ifx\testbx\f@series
{\mathbf{#1\,\mathsc{#2}}}\else
{\mathrm{#1\,\mathsc{#2}}}\fi
\else\textup{#1\,{\mdseries\textsc{#2}}}%
\fi}

\graphicspath{{Figures/}}

\begin{document}

\title{First Science with SAMI: A Serendipitously Discovered Galactic Wind in ESO 185-G031}

\author{Lisa M. R. Fogarty\altaffilmark{1}}
\author{Joss Bland-Hawthorn\altaffilmark{1}}
\author{Scott M. Croom\altaffilmark{1, 2}}
\author{Andrew W. Green\altaffilmark{3}}

\author{Julia J. Bryant\altaffilmark{1, 2}}
\author{Jon S. Lawrence\altaffilmark{3}}
\author{Samuel Richards\altaffilmark{1}}

\author{James T. Allen\altaffilmark{1}}
\author{Amanda E. Bauer\altaffilmark{3}}
\author{Michael N. Birchall\altaffilmark{3}}
\author{Sarah Brough\altaffilmark{3}}
\author{Matthew Colless\altaffilmark{3}}
\author{Simon C. Ellis\altaffilmark{3}} 
\author{Tony Farrell\altaffilmark{3}}
\author{Michael Goodwin\altaffilmark{3}}
\author{Ron Heald\altaffilmark{3}} 
\author{Andrew M. Hopkins\altaffilmark{3}} 
\author{Anthony Horton\altaffilmark{3}} 
\author{D. Heath Jones\altaffilmark{4}} 
\author{Steve Lee\altaffilmark{3}} 
\author{Geraint Lewis\altaffilmark{1}}
\author{\'Angel R. L\'opez-S\'anchez\altaffilmark{3, 5}} 
\author{Stan Miziarski\altaffilmark{3}} 
\author{Holly Trowland\altaffilmark{1}}

\author{Sergio G. Leon-Saval\altaffilmark{1}} 
\author{Seong-Sik Min\altaffilmark{1}} 
\author{Christopher Trinh\altaffilmark{1}} 

\author{Gerald Cecil$^{6}$}
\author{Sylvain Veilleux$^{7}$}
\author{Kory Kreimeyer$^{7}$}

\altaffiltext{1}{Sydney Institute for Astronomy (SIfA), School of
Physics, The University of Sydney, NSW 2006, Australia}
\altaffiltext{2}{ARC Centre of Excellence for All-sky Astrophysics (CAASTRO)}
\altaffiltext{3}{Australian Astronomical Observatory, PO Box 296, Epping, NSW 1710,
Australia}
\altaffiltext{4}{School of Physics, Monash University, Clayton, VIC 3800, Australia}
\altaffiltext{5}{Department of Physics \& Astronomy, Macquarie University, NSW 2109, Australia}
\altaffiltext{6}{Department of Physics \& Astronomy, University of North Carolina, Chapel Hill, NC, USA}
\altaffiltext{7}{Astronomy Program, University of Maryland, Baltimore, MD, USA}

\begin{abstract}
We present the first scientific results from the Sydney-AAO Multi-Object IFS (SAMI) 
at the Anglo-Australian Telescope. This unique instrument deploys 13 fused 
fibre bundles (hexabundles) across a one-degree field of view allowing simultaneous spatially-resolved spectroscopy of 13 galaxies. During the first SAMI commissioning run, targeting a single galaxy field, one object (ESO 185-G031) was found to have extended
minor axis emission with ionisation and kinematic properties consistent with a 
large-scale galactic wind. The importance of this result is two-fold: (i) 
fibre bundle spectrographs are able to identify low-surface brightness emission
arising from extranuclear activity; (ii) such activity may be more common than 
presently assumed because conventional multi-object spectrographs use single-aperture
fibres and spectra from these are nearly always dominated by nuclear emission. These early results
demonstrate the extraordinary potential of multi-object hexabundle spectroscopy in
future galaxy surveys.
\end{abstract}

\keywords{star formation, galaxy: kinematics, galaxy: ionisation, spectroscopy}

\section{Introduction}

\subsection{Feedback and Galactic Winds}

Feedback, from stars and active galactic nuclei (AGN), plays a vital role in regulating star formation in galaxies, thus greatly impacting their growth and evolution. One of the most spectacular feedback phenomena observed throughout the cosmos are galactic winds. Their effects can be seen across the electromagnetic spectrum from radio waves \citep{Duric83}, mid-infrared \citep{Roussel10}, and UV \citep{Hoopes05} through to X-rays \citep{Strickland04}, and even gamma rays \citep{Su10}. The importance of winds in galactic evolution is still unclear although they are suspected to play a role in the observed mass-metallicity relation \citep{DekelSilk86} and the reduced baryon fraction in galaxies compared to the universal background \citep{McGaugh10}. Outflows are routinely invoked in numerical simulations to forcibly remove the build-up of low angular momentum material at the centre of synthetic galaxies \citep{Bertone05, Brook10, Brooks10}. At the present time, this appears to be the only viable mechanism for producing exponential disks \citep{Sharma12}.

Large-scale winds are common in starburst and active galaxies and are therefore likely to be important at high redshifts where starbursts dominate star formation and AGN activity was at its peak \citep{Veilleux05}. In general, winds are highly energetic and largely invisible, where their presence is betrayed by material entrained from the galactic disk. This is evident from the slow rotation of outflowing filaments about the wind axis \citep{Shopbell98, Greve04}.

There has never been a systematic survey to assess the statistical occurrence of galactic wind sources in any waveband. While integrated spectra can show evidence of outflows, very detailed observations are required to properly separate the faint wind material from the background galaxy \citep{Cecil01}. The clearest indication of a wind typically comes from either (i) spatially resolved X-ray imaging \citep{Strickland04}; or (ii) combining spatially resolved ionization diagnostics and kinematic information \citep{Veilleux02, SBH10}, as we demonstrate here.

Because of the need for detailed spatially-resolved data, it has been difficult to envisage how a large-scale systematic survey of galactic winds would be possible if these amount to only 1\% of all galaxies \citep{Veilleux05}. With an instrument concept such as the Sydney-AAO Multi-object Integral field spectrograph (IFS) \citep[SAMI;][]{Croom12} a survey of this kind is now possible. As argued below, we envisage surveys of thousands of galaxies, where every object has spatially resolved IFS data, enabling the detection of extended, low surface brightness material and thus the identification of complex structures and spatial gradients in galaxy properties.

\subsection{The Sydney-AAO Multi-object Integral field spectrograph (SAMI)}

The Sydney-AAO Multi-object IFS is an innovative new instrument on the Anglo-Australian Telescope (AAT), providing 13 deployable fused fibre bundles, called hexabundles, over a large field of view. One of the strongest science motivations behind such an instrument is the study of complexity in galaxies. Understanding complex processes, in particular the flow of gas in and out of galaxies, is vital for solving the major problems of galaxy formation and evolution. Phenomena such as accretion, feedback, and the dynamical disturbance of gas through external and secular processes can manifest themselves in subtle ways via spatial gradients and/or structure in chemical abundances, ionisation, and kinematic properties, to name a few. Finding such sub-structure in galaxy properties is not trivial and relies on spatially-resolved spectral data. It is generally not possible for single-aperture spectroscopic studies to identify what kinds of objects exhibit these features. The discovery space probed with multiplexed IFS allows us to simultaneously {\it find} and study such objects in detail.

SAMI is the first large-scale multiplexed IFS to use hexabundles, a new optical fibre imaging bundle designed and developed at the University of Sydney \citep{JBH11, Bryant11}. Hexabundles consist of optical fibres with reduced cladding arranged in a circular pattern and lightly fused over a short length, yielding an imaging bundle with a high fill factor and excellent optical performance \citep{Bryant11}. SAMI uses 13 hexabundles deployable across a 1-degree field of view with each of the SAMI bundles containing 61 fibres. The instrument also contains 26 separate single fibres for sky subtraction. SAMI feeds the double-beam AAOmega spectrograph \citep{Sharp06}. 

SAMI is thus designed to simultaneously observe 13 spatially extended objects (e.g. galaxies), with 61 spatial samples across each target and spectral information for each of these spatial samples. The power of this technique is immediately apparent if we consider two things - the aperture effects inherent in single fibre galaxy surveys and the time-consuming nature of current IFS surveys. 

The effects of a single fixed aperture size (e.g. a fibre on sky) on large galaxy surveys have been known for some time. The problem is simple: the properties derived from aperture spectra of galaxies depend on both the position and size of the aperture one is observing with. For galaxies larger than the aperture only the central part of the galaxy is observed and some of the light in the outskirts of the object is lost. Since the apparent size of a galaxy on sky will vary with redshift, even over modest redshift ranges, this makes it very difficult to compare samples across different epochs and could affect the interpretation of the evolution of global galaxy properties. Likewise, if the observing aperture is not well centred on a galaxy it is difficult to say what relevance the derived properties will have to the object's true integrated properties. This is particularly relevant if a galaxy is not morphologically well-behaved, for example, how is one to determine the ``centre" position of a merging or otherwise disturbed system? 

This problem is addressed in various ways using complex aperture corrections (e.g. \citet{Brinchmann04} for SDSS). While these aperture corrections can be shown to be broadly correct when applied to a population of galaxies (albeit with significant scatter remaining in the derived properties, see \citet{Hopkins03}), they can suffer gross inaccuracies when dealing with individual galaxies. For example \citet{Gerssen12} found that for a sample of star-forming galaxies in SDSS, the star formation rates derived by \citet{Brinchmann04} were underestimated by a median factor of 2.5, with a scatter of 175\%. As well as this need for aperture correction, single fibre spectra are inherently limited as they can only reveal the {\it average} properties of galaxies. In contrast to IFS data, any spatial variation across the object itself is lost.

Although the need for them is clear, large surveys of galaxies with IFS data have heretofore been difficult and time-consuming to execute. This is because most IFS are single-object instruments, meaning only one galaxy can be observed at a time. The largest sample to date is that of the ATLAS-3D project \citep{ATLAS-3D}, numbering 260 early-type galaxies each observed with the SAURON spectrograph. The CALIFA project \citep{Sanchez12} aims to build a sample of 600 galaxies using the PMAS IFS. At higher redshift the IMAGES survey \citep{Yang08} used the FLAMES/GIRAFFE MOS-IFU system to observe a sample of 63 galaxies at redshifts z$\sim$0.4-0.75, while the SINS survey \citep{ForsterSchreiber09} used the SINFONI IFS to observe a sample of 63 galaxies at redshifts z$\sim$1.2-3.6. These samples clearly demonstrate the richness and potential of large IFS samples but were extremely time-consuming to observe. With its multiplex factor of 13, SAMI will be able to observe a sample of $\sim$350 galaxies in twenty nights.

The first SAMI commissioning results were published in \citet{Croom12}. To further demonstrate the scientific capability of SAMI, here we take a closer look at just one of the objects observed during commissioning in July 2011. Specifically we present observations of a serendipitously discovered  galactic wind, whose detection was enabled by spatially resolved SAMI spectroscopy of the galaxy in question. This discovery demonstrates the power of multi-object IFS to identify rare objects that are impossible to detect in single integrated spectra.

Section \ref{sec:obs} describes our instrument set-up and observations, while Section \ref{sec:reds} describes our data reduction procedure with particular attention paid to the dithered SAMI observation. Section \ref{sec:res} describes our results, with Section \ref{sec:conc} laying out our conclusions. Our analysis uses the cosmological parameters provided by \citet{Komatsu09} with $\Omega_{\Lambda}$=0.723, $\Omega_{M}$=0.277 and H$_{0}$=70.2\,km\,s$^{-1}$Mpc$^{-1}$.

\section{Observations}
\label{sec:obs}

\begin{table}
\centering
\begin{tabular}{|c|c|c|c|}
\hline
Date & Run \# & Total Exposure Time (s) & Format \\ 
\hline
2nd July & 29-31 & 7200 & Pointed \\
\hline
3rd July & 29-37 & 14034 & Dithered \\
\hline
\end{tabular}
\caption{SAMI observations of the wind galaxy.}
\label{tab:obs}
\end{table}

\begin{figure}
\centering
\includegraphics[width=8.5cm]{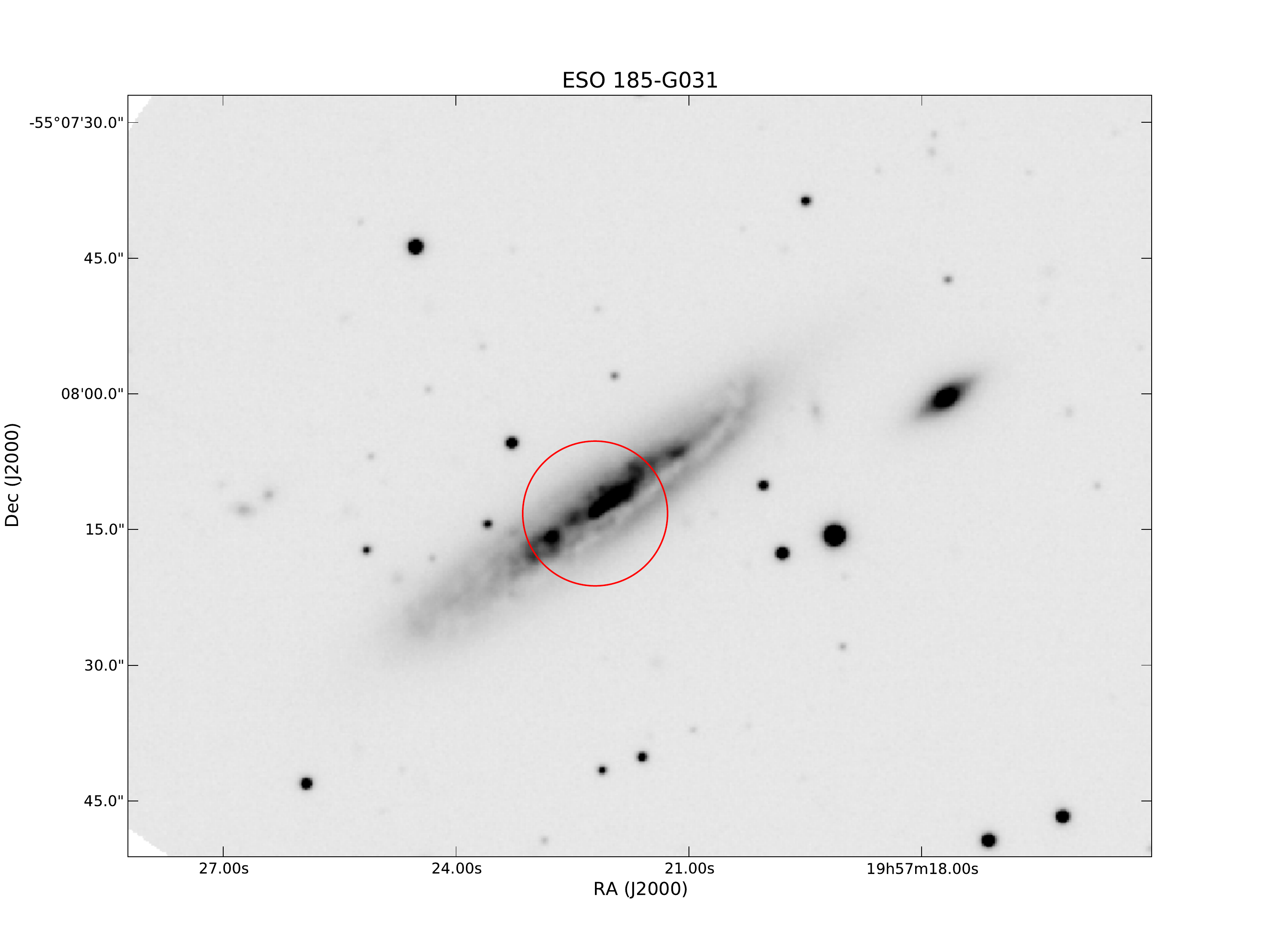}
\caption{An R-band image of ESO 185-G031 taken with the Maryland-Magellan Tunable Filter (MMTF) instrument on the 6.5-meter Walter Baade Magellan telescope at Las Campanas Observatory. The galaxy is a gas-rich spiral with well-defined arms. The near-side arm is clearly visible running south-east to north-west along the disk. The red circle indicates the footprint of the SAMI hexabundle on the galaxy.
\label{fig:im_mmtf}}
\end{figure}

SAMI was mounted at the prime focus of the AAT behind the triplet corrector which provides a 1-degree field of view with a plate scale of 15.2\arcsec/mm. Each of the SAMI hexabundles contains 61 105$\mu$m-diameter fibres in a circular pattern, meaning each hexabundle has a field of view of 14\farcs{9}\, on sky, with a sampling of 1\farcs{6} per fibre. From its position at prime focus, a 42m fibre cable runs from SAMI to the AAOmega spectrograph in the coud\'{e} room.

AAOmega is a double-beam spectrograph which can operate with a range of dichroics and gratings, yielding a corresponding range of values for resolution and wavelength coverage in each spectrograph arm. For these observations we used the 580V grating in the blue arm, providing spectral coverage from 3700-5700\,\AA\, and a resolution of $R=\lambda/\Delta\lambda\sim1730$. This grating was chosen to cover the most interesting spectral features in the blue end of the spectrum, including the [\ion{O}{ii}]~$\lambda$\,3727 \AA\, emission feature and the ``4000-\r{A}ngstrom" continuum break, $D_{4000}$. However, since the type of fibre used for the 42m-long SAMI fibre run was shown to have poor throughput in the blue end of the spectrum (this fibre run is currently being upgraded, see \citet{Croom12} for more details) we did not observe any usable data below 4000\,\AA, with the throughput beginning to decline for $\lambda<4500$\,\AA. We did, however, robustly measure other features redward of 4500\,\AA, including, for this galaxy, both the H$\beta$ and [\ion{O}{iii}]~$\lambda$$\lambda$\,4959, 5007\,\AA\, doublet emission features.

In the red arm, we used the 1000R grating providing spectral coverage over the range 6250-7350\,\AA\, at a resolution of $R\sim4500$. This grating was chosen to provide high resolution measurements of the H$\alpha$ line for the derivation of kinematic properties. The wavelength range also covers the [\ion{O}{i}]~$\lambda$\,6300, [\ion{N}{ii}]~$\lambda$\,6583 and [\ion{S}{ii}]~$\lambda$$\lambda$\,6717, 6731\,\AA\, emission features.

For the first SAMI commissioning run in July 2011, one field containing 13 galaxies was observed. The galaxies were selected from the 6dF Galaxy Survey \citep[6dFGS;][]{Jones04} to have sufficient source density for SAMI, be large enough to fill the spatial extent of the hexabundles (or, indeed, overfill them) and to have a surface brightness sufficient to give S/N$\geq$3 in the outermost fibres in a 2-hour exposure. The reader is referred to Section 4.2 of \citet{Croom12} for more details on the target selection.

The observations are summarised in Table \ref{tab:obs}. The data were observed using one of two strategies, either pointed or dithered, with the latter strategy being used to increase the spatial sampling and coverage of the observations. The pointed data were observed on the 2nd July with a cloudless sky. The dithered data were observed on the 3rd July in less optimal conditions. For this paper the pointed and dithered data are kept separate for comparison. 

For the remainder of this paper, we concentrate on a single galaxy from the sample of 13 which were observed. This galaxy has 6dFGS identification g1957222-550814 and ESO identification ESO 185-G031 and is at a redshift of z=0.016. This corresponds to an 
approximate distance of 68 Mpc and an angular scale of 1$\arcsec$ equivalent to 
340 pc. An R-band image of the galaxy is shown in Figure \ref{fig:im_mmtf}.

\subsection{Imaging}

In addition to the SAMI data, ESO 185-G031 was observed with the Maryland-Magellan Tunable Filter (MMTF) and the f/2 camera of the Inamori-Magellan Areal Camera and Spectrograph (IMACS) camera on the 6.5-meter Walter Baade Magellan telescope at Las Campanas Observatory on 22nd August 2011. The galaxy was observed with a Bessel R-band filter for three one-minute exposures. The MMTF order-blocking filter was then used at a central wavelength of 6660\,\AA\, and a bandpass of 260\,\AA. Four ten-minute exposures were taken in this configuration, yielding a total of 40 minutes on source.

The combined three-minute R-band image of the galaxy is shown in Figure \ref{fig:im_mmtf}. The large red circle shows the position of the SAMI hexabundle for the pointed data. The dithers are not shown as they were small (of the order of 1/2 core diameter, or 0\farcs{8}) and there were nine individual pointings observed. The smaller green circle, with a radius of 1\arcsec\,shows the position in the galaxy of the supernova SN1998bo, a type Ic supernova seen in 1998. The data reduction and analysis are discussed in Sections \ref{sec:im_reds} and \ref{sec:im_analysis} respectively.

\section{Data Reduction}
\label{sec:reds}

The raw SAMI data were reduced using the 2{\sc dfdr} data reduction package \citep{CSH04, SharpBirchall10}, a robust package already used for both 2dF and SPIRAL data and expanded by our team to process SAMI data. It performs bias subtraction and flat-fielding before extracting the individual fibre spectra. The extracted spectra are then wavelength calibrated, sky subtracted and corrected for fibre-to-fibre throughput variations, before being reformatted into a row-stacked spectrum (RSS) frame consisting of all 819 extracted SAMI spectra (61 spectra for each of the 13 hexabundles plus 26 separate sky spectra). 

\begin{figure*}
\centering
\includegraphics[width=\textwidth]{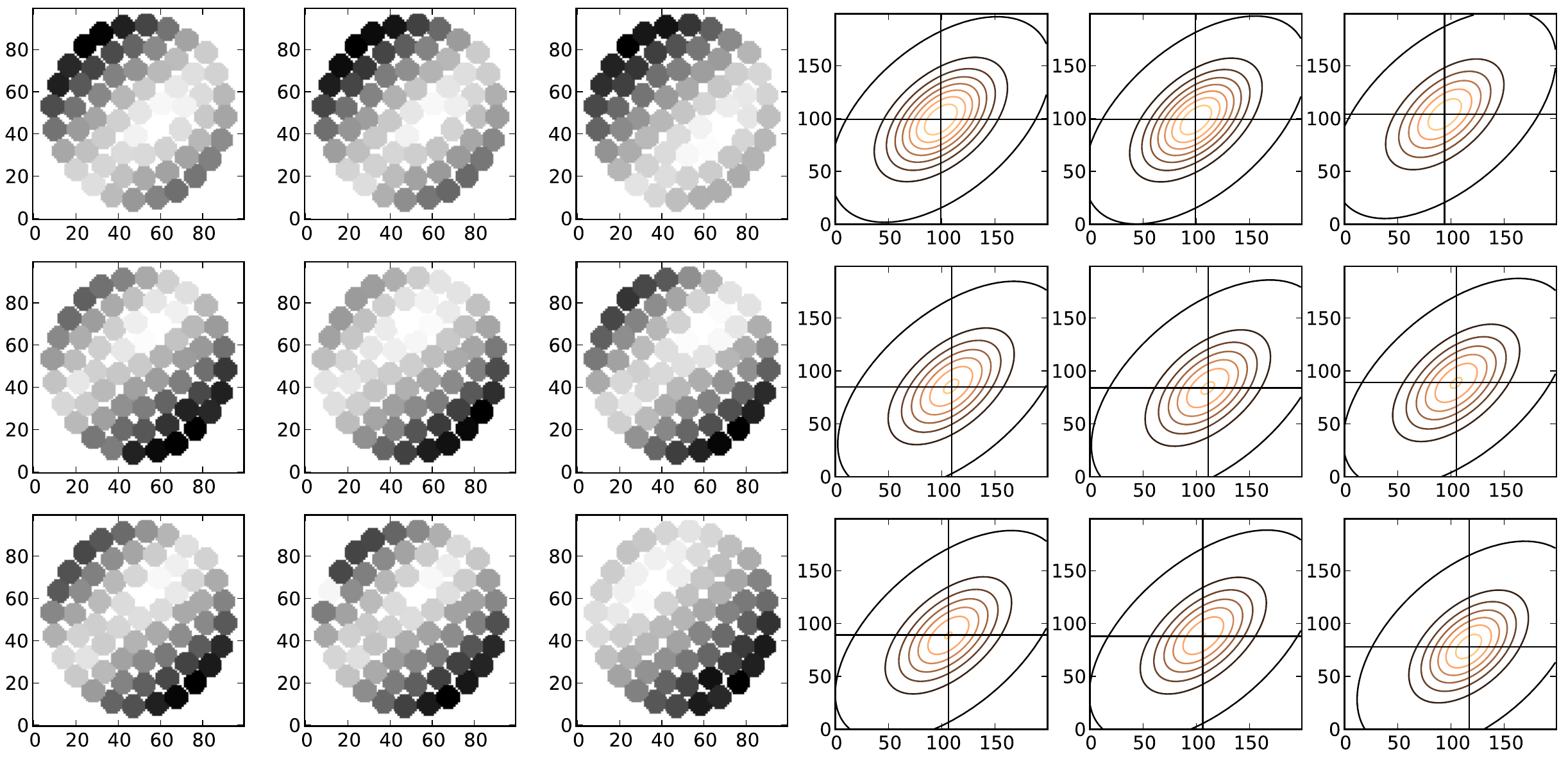}
\caption{Left-hand panel: Integrated images of the sequence of dithered data cubes observed on 3rd July 2011. The images were formed by collapsing each data cube along the wavelength axis from 6300\,\AA-7300\,\AA. The shift from one observation to the next is clearly seen. To combine these data cubes it is necessary to calculate the positional offsets between each observation. Right-hand panel: The contours show the 2d Gaussian fit to the cross-correlation of each of the images in the left-hand panel with the fiducial image (in this case this is the first image in the sequence of dithers, so the first contour plot shown here is an auto-correlation). The postional offsets between the images are calculated from these fits.
\label{fig:ims_xcorr}}
\end{figure*}

For this data set, a spectrophotometric standard star, LTT6248, was observed on 2nd July. The star was observed in one hexabundle only and was reduced in exactly the same manner as the data frames. A ``total'' measured star spectrum is found by summing the extracted SAMI spectra. The measured star spectrum is compared to the tabulated flux values for the star and a sensitivity function found for the SAMI spectra. This is normalised and applied to the extracted spectra, yielding relative flux-calibrated RSS frames for each SAMI observation. The final step in the reduction is to scale each frame by the exposure time to be in units of counts/second, thus making all frames directly comparable. 

The three pointed frames observed on 2nd July were then combined yielding a single RSS frame with a 2 hour exposure time and containing all thirteen galaxies in the field.

\subsection{The Dithered Data}

The SAMI hexabundles consist of a circular pattern of circular optical fibres. By its nature, the fibre coverage is not contiguous and, although the SAMI hexabundles have a high filling factor (greater than 75\%), there are inevitably some gaps between the fibres. In addition, the circular fibres are not stacked in close-packed configuration, but in a circular pattern, meaning the sampling also forms a non-uniform grid. These details present certain challenges when working with dithered data. The non-contiguous, non-uniform coverage makes it difficult to simply interpolate over the holes in the sampling. We experimented with different ways to process the dithered data and our most consistent technique is described below.

For the dithered data RSS frames were produced as usual using 2{\sc dfdr} but before combining we first reconstruct entire data cubes for ESO 185-G031 from each SAMI observation.
The SAMI data cubes used in this paper are created using a supersampled position grid at each wavelength slice. The sampling of the new grid is at the level of 1/10th of a core diameter (0\farcs{16}). The position of the centre of each core is defined to be the pixel closest to the tabulated position of that core, and so is determined on the new grid to an accuracy of 1/10th of a core. A core is then defined to lie at that position in all wavelength slices. The core is ``drawn'' as an approximate circle and filled with the spectrum recorded for that core. In this way each core is represented by a pixellated circle on the spatial grid. This represents the spatial extent of each sample without interpolating the data at this early stage.

Once all data cubes are in hand they must be combined, accurately accounting for the positional offsets between the cubes. To calculate the offsets between each cube in pixel space a cross-correlation method was used. First, each cube was collapsed over a large wavelength range (6300\,-\,7300\,\AA) to create an integrated image for each pointing. These images were then cross-correlated against a fiducial image, in this case simply the first SAMI image in our observing sequence. The resulting correlation function was then fit with a two-dimensional Gaussian to find the position of peak and therefore the offset from the peak position in the fiducial image. This allowed the data cubes to be accurately combined at the pixel level. The SAMI images and fitted cross-correlation functions are shown in the left-hand and right-hand panels of Figure \ref{fig:ims_xcorr} respectively. 

The cubes were combined by shifting and median stacking them using the offsets in pixel space from the cross-correlation method. This yielded a single combined dithered data cube. Note that the depth and therefore S/N of the dithered observation will vary across the field of view, again because of the non-contiguous sampling inherent in the hexabundles.

For the analysis presented in this paper the combined cubes were then smoothed spatially with a Gaussian kernel with a 5-pixel sigma and spatially re-binned by a factor of both two and four, yielding four final data cubes, two finely sampled (red and blue) and two more coarsely sampled (red and blue). The more coarsely sampled cubes are necessary for sufficient S/N per spatial pixel in the blue and unless otherwise stated these are used for all further analysis.

The effects of differential atmospheric refraction have not been corrected for in the final data cubes. However, following the formulation in \citet{Filippenko82} the effect is found to be small and does not impact the results of this paper. Within the individual line ratios the wavelength ranges are very short and the effect is negligible. Over the entire wavelength range considered when comparing line ratios with one another (i.e. 4900\,-6900\,\AA) the relative shift is 0.35\arcsec. This is half a pixel in the coarsely binned SAMI cubes (upon which this analysis was performed) and will therefore not affect our results.

\subsection{Imaging}
\label{sec:im_reds}

The MMTF data reduction pipeline was used to reduce the imaging data from IMMTF/MACS. Bias subtraction and flat-fielding were performed using Image Reduction and Analysis Facility ({\sc iraf}) scripts and cosmic rays were identified and rejected using the {\sc LACosmic} routine of  \citet{vanDokkum01}. A median sky value for the entire image was calculated and subtracted. The astrometry for each image was found using the standard distortion map for IMACS, with some small adjustments. The images were registered using the USNO catalogue. The same procedure was applied to both the line and continuum images.

\section{Results}
\label{sec:res}

\begin{figure*}[!ht]
\centering
\includegraphics[width=\textwidth]{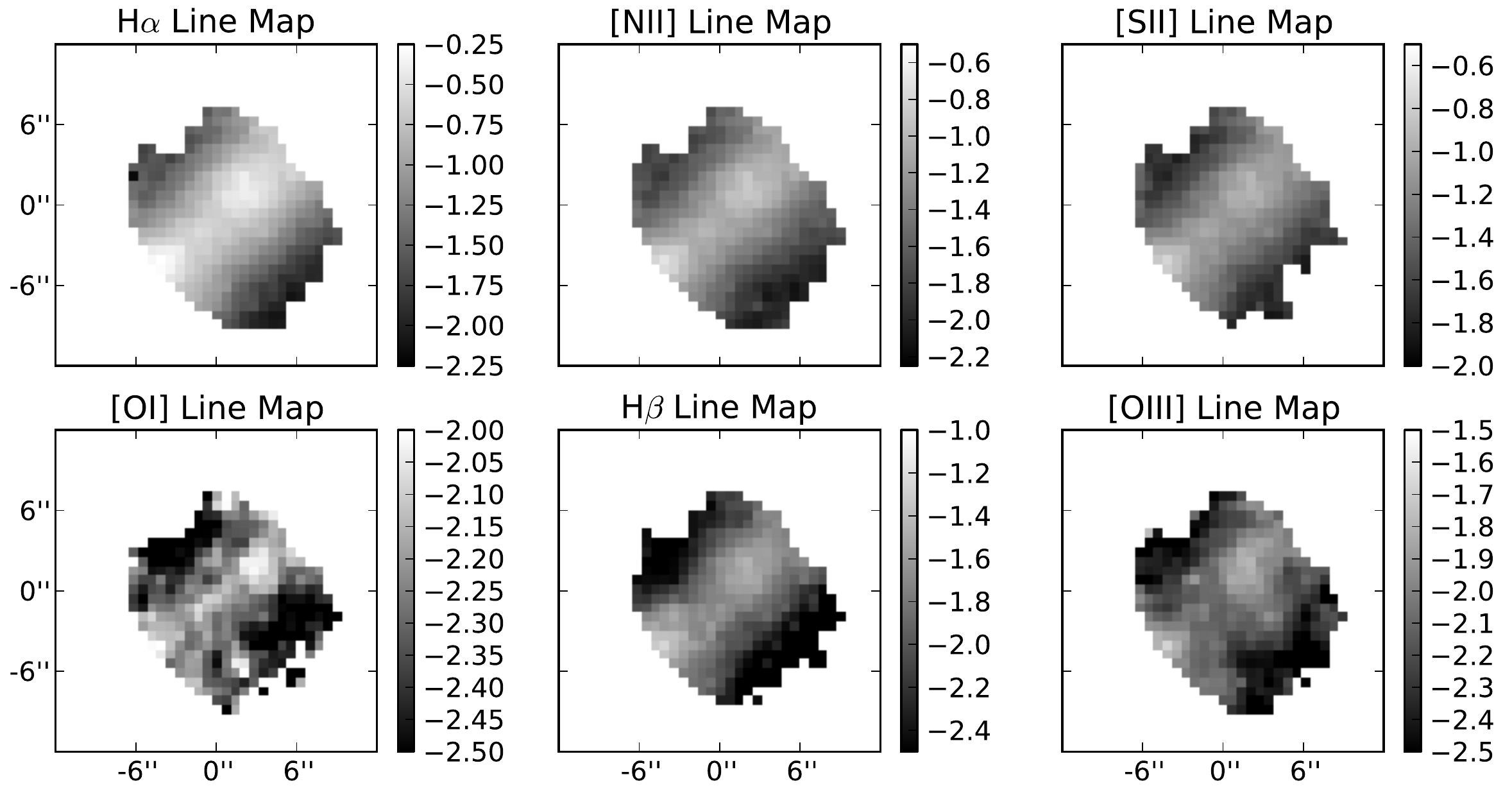}
\caption{Spatial maps of the emission line strengths derived from the dithered data where the [\ion{S}{ii}] map is a sum of the two lines in the doublet. The disk is clearly seen in emission in all lines. The colour bar is in units of log$_{10}$(counts/s). The images are 24\arcsec on a side.
\label{fig:linemaps}}
\end{figure*}

\begin{figure*}[!ht]
\centering
\includegraphics[width=\textwidth]{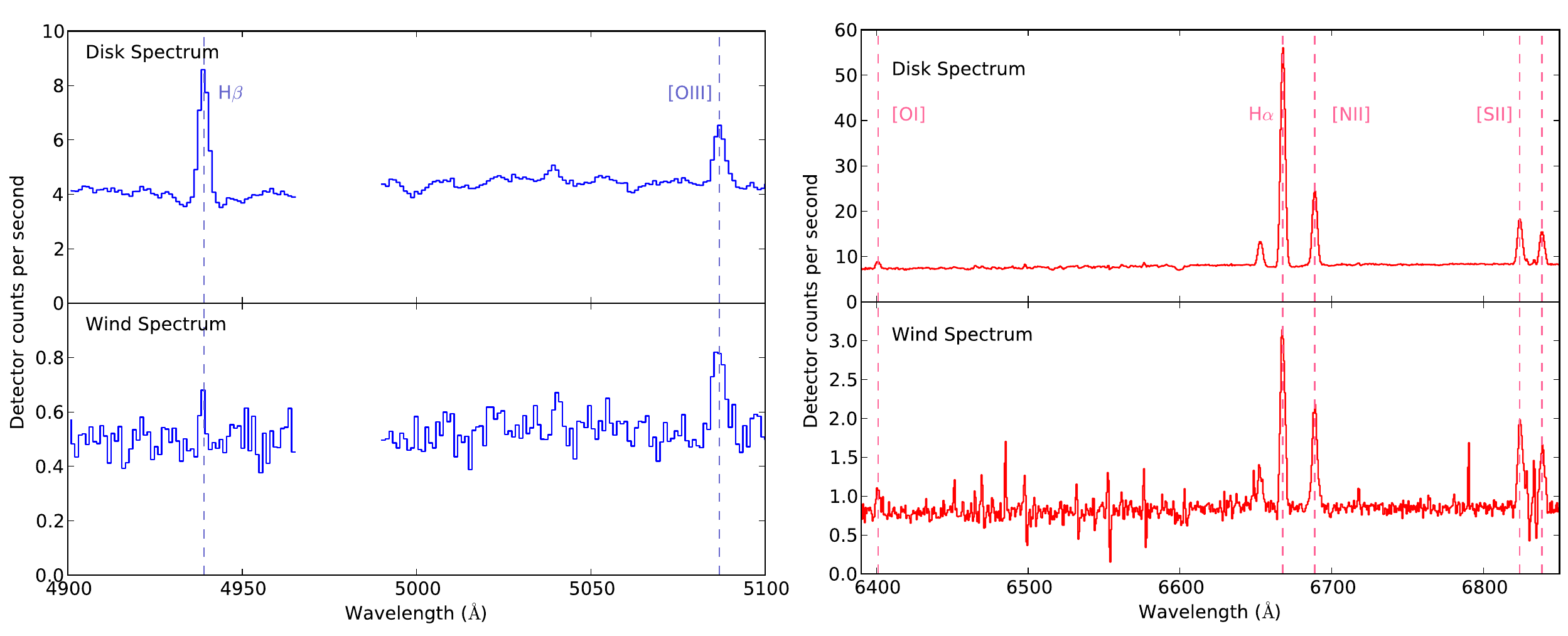}
\caption{Left: two integrated SAMI spectra, from the disk and wind regions of ESO 185-G031, covering the blue wavelength range from 4900\,-5100\,\AA. The H$\beta$ and [\ion{O}{iii}] emission lines are clearly seen in both spectra, but with lower S/N in the wind region. Right: two integrated SAMI spectra, from the disk and wind regions of ESO 185-G031, covering the red wavelength range from 6400\,-6850\,\AA. The [\ion{O}{i}], [\ion{N}{ii}], H$\alpha$ and [\ion{S}{ii}] emission lines are clearly seen in both spectra, but with lower S/N in the wind region, where the sky residuals are also more prominent.
\label{fig:spectra_tot}}
\end{figure*}

\begin{figure}
\centering
\includegraphics[width=8.0cm]{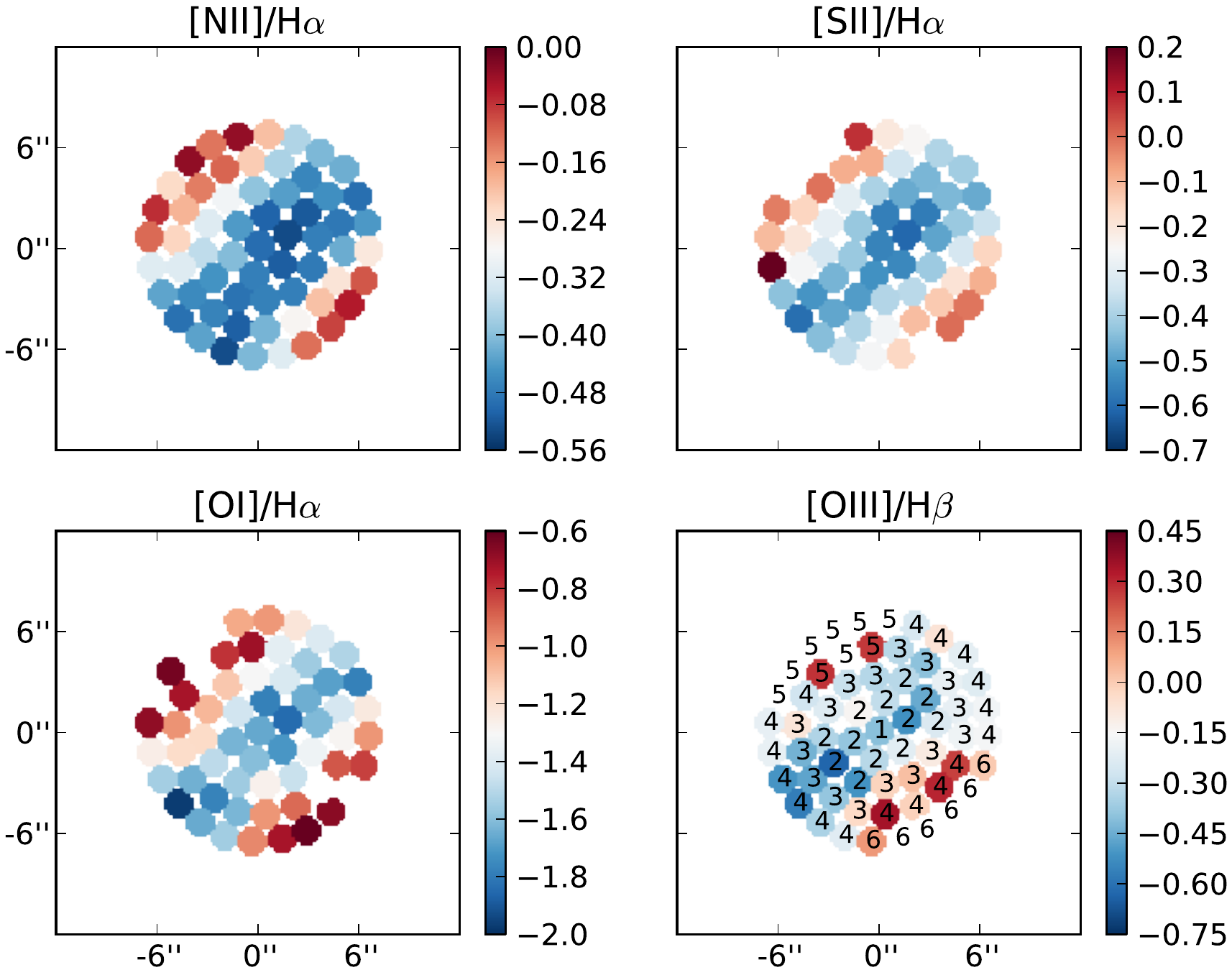}
\caption{Spatial maps of the emission line ratios derived from the pointed data. Blank spaxels correspond to spectra where the S/N in the emission lines in question was not high enough to obtain a good fit. These maps are used to construct the IDDs in Figure \ref{fig:bpt_2hr}. Since the [\ion{O}{iii}]/H$\beta$ ratio (final panel) is needed for all three IDDs, incompleteness in this map clearly has the greatest impact. The final panel shows the spaxels numbered in groups used to extract aperture spectra (see Section \ref{sec:idd_pointed} and Figure \ref{fig:bpt_2hr}). In all maps a trend towards higher line ratios is seen off the plane of the disk of the galaxy, indicating the presence of a galactic wind. The images are 24\arcsec on a side.
\label{fig:ratios_2hr}}
\end{figure}

\begin{figure}
\centering
\includegraphics[width=8.0cm]{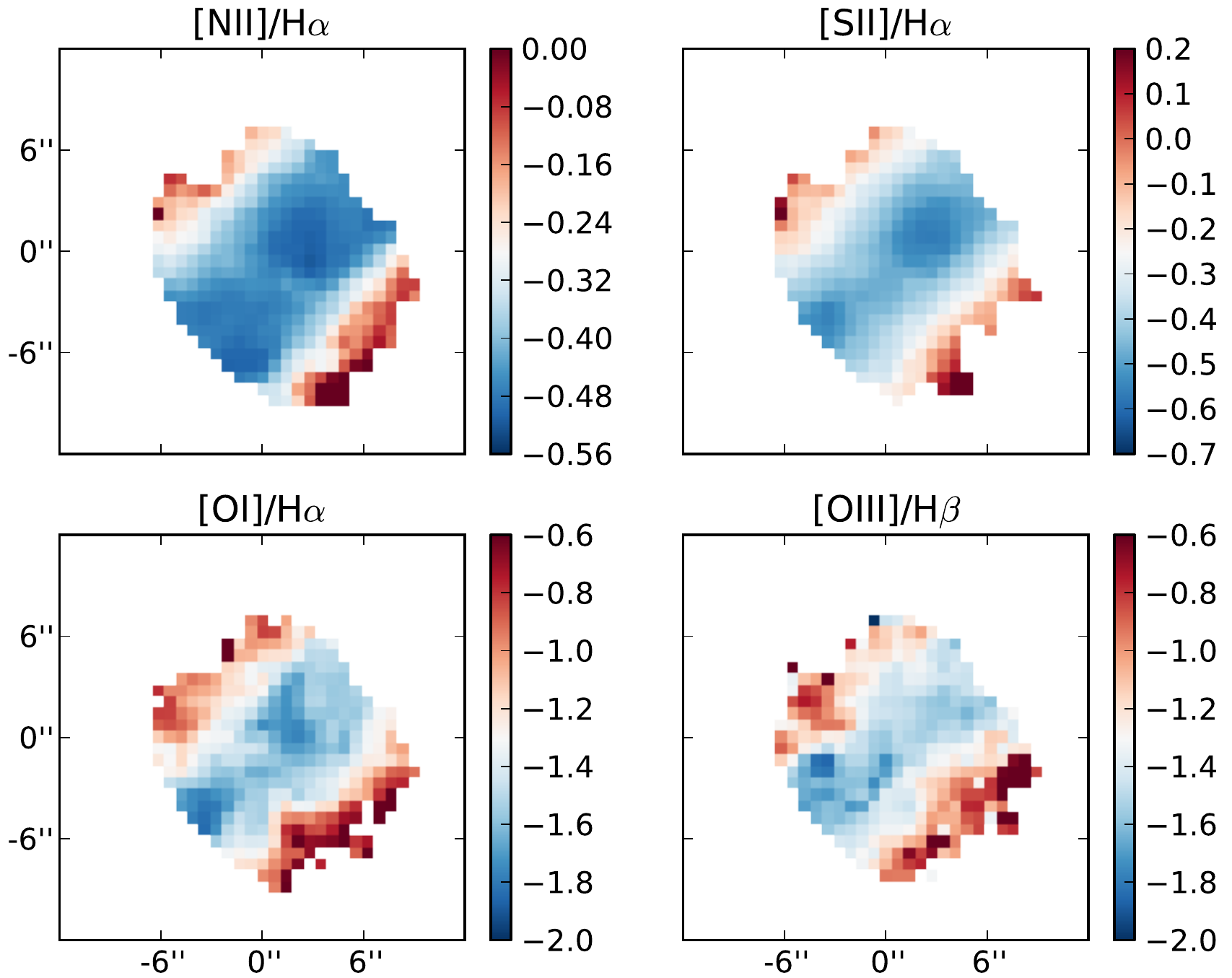}
\caption{Spatial maps of the emission line ratios derived from the dithered data. Again, the blank spaxels are those without spatial coverage or sufficient S/N to fit the necessary lines. These are used to construct the IDDs in Figure \ref{fig:bpt_dith}. The trend towards higher line ratios off the plane of the disk is clearly seen, indicating the presence of a galactic wind. The images are 24\arcsec on a side.
\label{fig:ratios_dith}}
\end{figure}

Once the SAMI data were reduced the analysis was performed using custom-written {\sc python} scripts. The method was the same for both the pointed and dithered data. For each spectrum seven individual emission lines were each fit with a Gaussian function. These were the H$\beta$, [\ion{O}{iii}]~$\lambda$\,5007, [\ion{O}{i}]~$\lambda$\,6300, H$\alpha$, [\ion{N}{ii}]~$\lambda$\,6583, [\ion{S}{ii}]~$\lambda$\,6717 and [\ion{S}{ii}]~$\lambda$\,6731\,\AA\, lines. Where it was detected the underlying stellar H$\beta$ absorption was also fit with a Gaussian and removed before fitting the emission line. Line strengths and kinematic properties were derived from the fits to each line. Six line strength maps derived from the dithered data are shown in Figure \ref{fig:linemaps}. Note that to produce the [\ion{S}{ii}] map the line strengths of the two doublet lines were summed. From these six line maps it is possible to also derive four common emission line ratios (in logarithmic units), these being [\ion{O}{iii}]/H$\beta$, [\ion{O}{i}]/H$\alpha$, [\ion{N}{ii}]/H$\alpha$ and [\ion{S}{ii}]/H$\alpha$. The spatial maps of these four ratios are shown in Figures \ref{fig:ratios_2hr} and \ref{fig:ratios_dith}, for the pointed and dithered data respectively. The [\ion{N}{ii}]/H$\alpha$ line ratio map is used to define spatial regions in the galaxy which are dominated by the disk  and the galactic wind, using a threshold of [\ion{N}{ii}]/H$\alpha>-0.35$ to isolate regions of the galaxy showing only ionisation due to star formation. Resulting integrated spectra are shown in Figure \ref{fig:spectra_tot}.

 \subsection{Ionisation Diagnostic Diagrams}
 
First proposed by \citet{BPT81} and revised by \citet{VO87}, Ionisation Diagnostic Diagrams (IDDs) are used to classify the dominant mechanism of gas ionisation in galaxies using the ratios of commonly observed optical emission lines. The most frequently used of these diagrams are those three using the [\ion{O}{iii}]/H$\beta$ line ratio versus each of the [\ion{O}{i}]/H$\alpha$, [\ion{N}{ii}]/H$\alpha$ and [\ion{S}{ii}]/H$\alpha$ line ratios.

\begin{figure*}[!t]
\centering
\includegraphics[width=\textwidth]{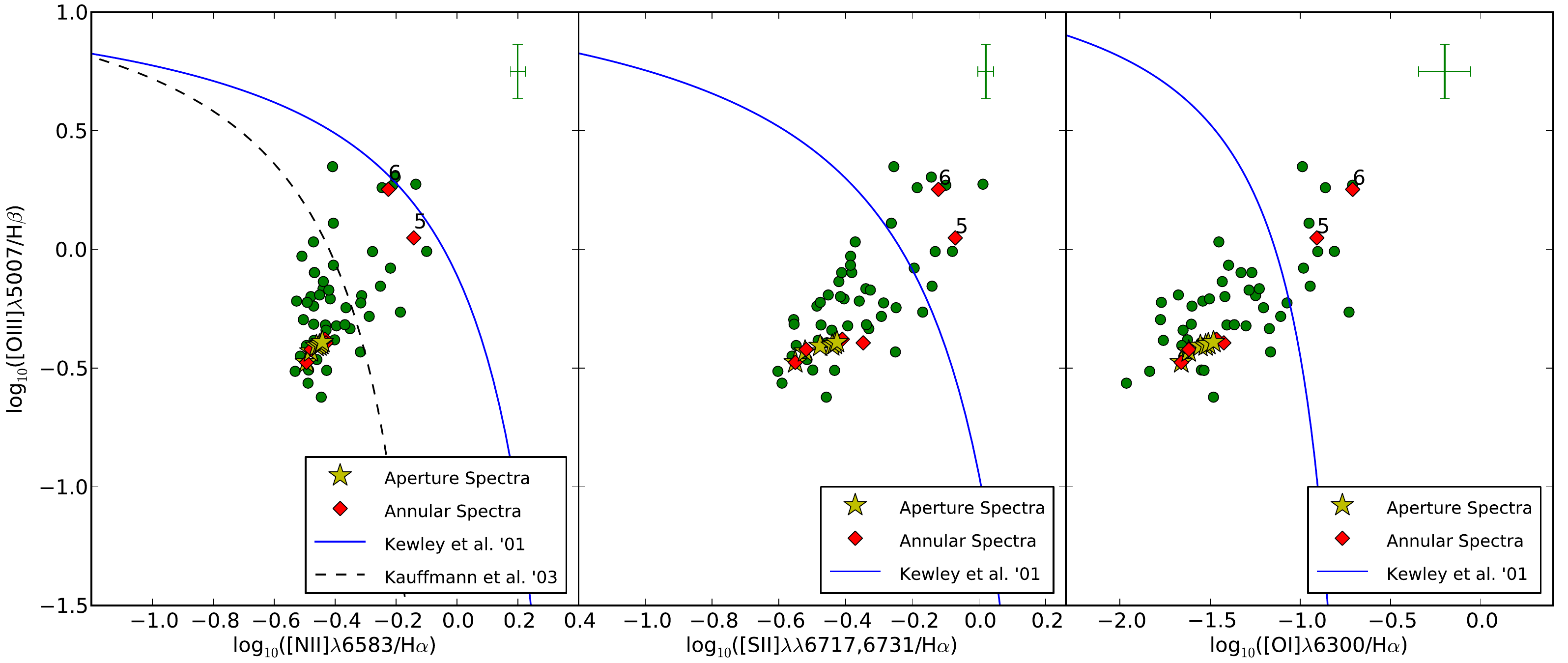}
\caption{IDDs derived from the pointed observations of ESO 185-G031. The \citet{Kewley01} limiting starburts curves are shown for all three IDDs (solid blue curve) with the \citet{Kauffmann03} delimiting line also shown for the [\ion{N}{ii}] IDD (dashed black curve). The yellow stars show the position of the galaxy for incrementally summed apertures, all of which indicate a system entirely dominated by star formation. The red diamonds show the same apertures treated as incremental annuli, with the outer two annuli labeled (5 and 6). This time a trend is seen such that the outer parts of the galaxy (annuli 5 and 6) show ionisation dominated by shocks from the newly discovered galactic wind. This clearly demonstrates the need for spatially-reolved spectroscopy to identify such systems in the future. A representative error bar is shown in green in the top-right corner of each plot.
\label{fig:bpt_2hr}}
\end{figure*}

\citet{Kewley01} used stellar population synthesis models as well as photoionisation models to define a theoretical maximum curve for star-forming objects on these three IDDs. Spectra of objects above this line (shown as the solid blue curves in our IDDs in Figures \ref{fig:bpt_2hr} and \ref{fig:bpt_dith}) are not dominated by star formation, but by a different source of ionisation, for example AGN or LINER emission. 

For the [\ion{N}{ii}] IDD, \citet{Kauffmann03} derived an alternative dividing line, based on empirical evidence from a large sample of galaxies. The \citet{Kauffmann03} curve (shown as the dashed black curve in our [\ion{N}{ii}] IDDs in Figures \ref{fig:bpt_2hr} and \ref{fig:bpt_dith}) separates purely star-forming objects (below the \citet{Kauffmann03} curve) from composite objects containing both star formation and AGN/LINER-like activity \citep{Kewley06}. Since we wish to separate the signature of star formation from any hint of other ionisation mechanisms (even if this other mechanism is not dominant) we adopt the \citet{Kauffmann03} line as the boundary between star formation and AGN/LINER-like emission in the [\ion{N}{ii}] IDD.

\subsubsection{The Pointed Data}
\label{sec:idd_pointed}
The IDDs for the pointed observations of ESO 185-G031 are shown in Figure \ref{fig:bpt_2hr}. The green points correspond to  individual spaxels for which both line ratios in question could be derived. (For some spaxels the S/N in the [\ion{O}{iii}] and H$\beta$ emission lines was too low to get a good fit. Since this ratio is required for all three IDDs, this is clearly the limiting factor here. See the final panel in Figure \ref{fig:ratios_2hr}.) 

For the pointed data summed aperture spectra are also considered and plotted as yellow stars in Figure \ref{fig:bpt_2hr}. The apertures are defined starting from the inner fibre and moving outwards, accounting for whole spaxels only, and roughly following the distribution of the [\ion{O}{iii}]/H$\beta$ emission line ratio (see the final panel in Figure \ref{fig:ratios_2hr}, where the numbers indicate spaxels summed to create the six aperture and annular spectra). The aperture spectra are found by summing the spectra within the aperture and the line ratios are derived as usual. As can be seen in Figure \ref{fig:bpt_2hr} all integrated aperture spectra (yellow stars) lie very close to or on top of one another. This shows that the light from the star forming disk is so dominant that integrated spectra over a fixed aperture, even if that aperture was the size of the entire SAMI hexabundle, would not pick up the non-star-forming signature we see off the plane of the disk.

The red diamonds in Figure \ref{fig:bpt_2hr} show the same apertures but now considered as annuli, i.e. this time each annulus is plotted individually instead of summing over all light within the aperture in question. This time a clear trend is seen, with the central annuli moving along a track through the star-forming region of the diagram towards the limiting curves and the outer two spectra clearly exhibiting properties indicative of a non-star-formation ionisation source. 

Again, it is clear that with a single fibre or a single large aperture these characteristics would have been missed. It is the spatial sampling of our data that allows us to discover the true nature of this galaxy.

\subsubsection{The Dithered Data}
\label{sec:IDD_dither}

\begin{figure*}
\centering
\includegraphics[width=\textwidth]{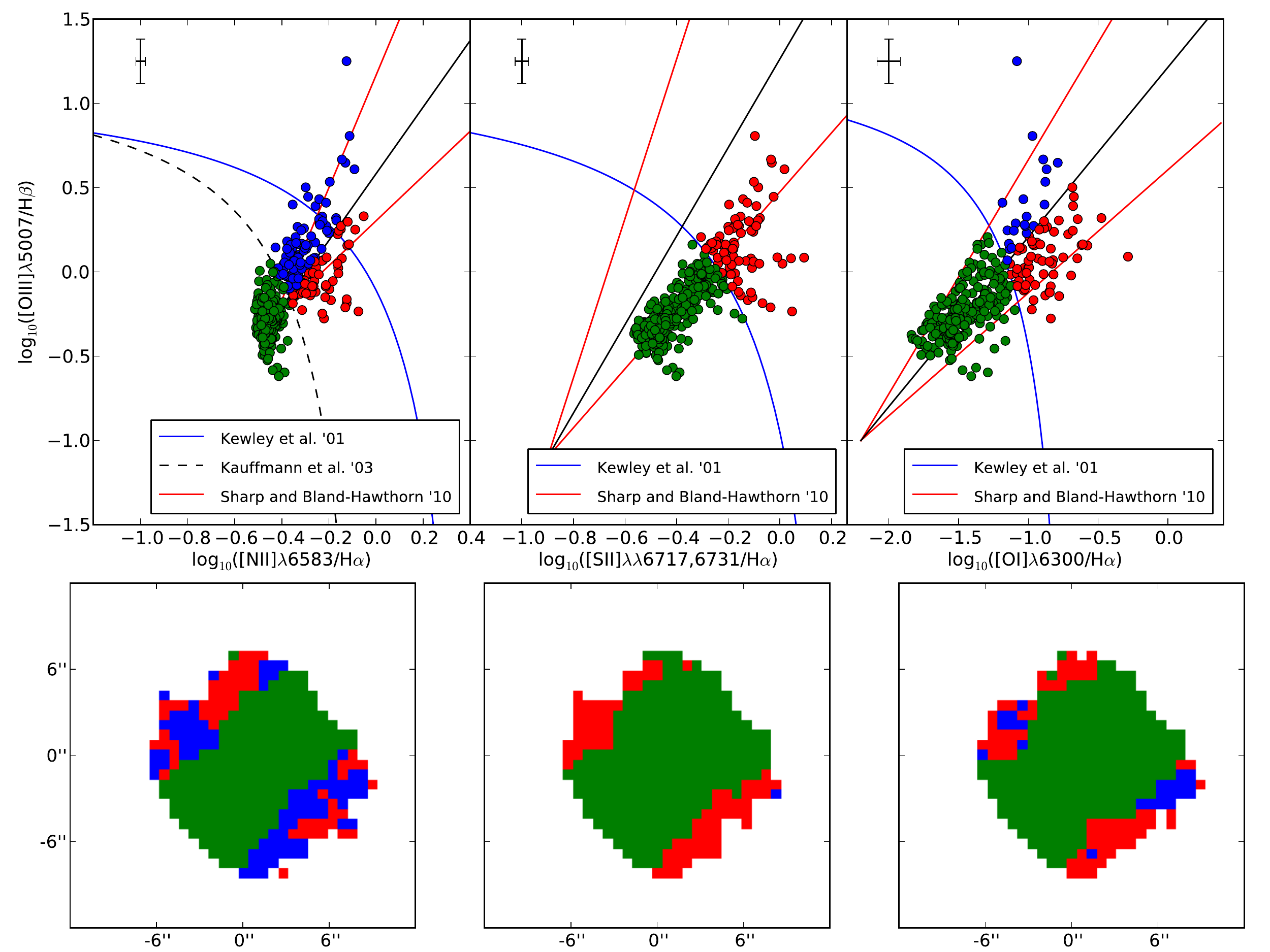}
\caption{Top panels: IDDs derived from the dithered observations of ESO 185-G031. Again, the \citet{Kewley01} limiting starburts curves are shown for all three IDDs (solid blue curve) with the \citet{Kauffmann03} delimiting line also shown for the [\ion{N}{ii}] IDD (dashed black curve). The colour coding is that of \citet{SBH10}, where green denotes spatial points dominated by star formation, blue denotes a dominant AGN-like emission and red denotes shock emission characteristic of a starburst driven wind. A representative error bar is shown in black in the top-left corner of each plot. Bottom panels: The colour-coded points are re-mapped to their spatial positions in the galaxy. All three maps show a star-formation-dominated disk with a stronger ionisation mechanism dominating off the disk in both directions. This is strong evidence for a galactic wind in this galaxy. See Section \ref{sec:IDD_dither} for more details. 
\label{fig:bpt_dith}}
\end{figure*} 

The IDDs for the dithered data are shown in the top part of Figure \ref{fig:bpt_dith}. As well as the \citet{Kewley01} and \citet{Kauffmann03} lines discussed above, we also present two fiducial lines following the work of \citet{SBH10}. These are shown as solid red lines in all three IDDs. The shallower line is a fit to the track followed by points in NGC1482 with increasing height above the plane of the disk of the galaxy. Since NGC1482 is a canonical starburst-driven galactic wind \citep{Veilleux02}, this line is taken to represent the trend expected in objects of that kind. The points lie in this part of the diagram because the ionisation source is shock excitation caused by the starburst-driven wind in the galaxy. Likewise, the steeper line is a fit to the track followed by points in NGC1365, a canonical AGN-driven wind \citep{Veron80} and is taken to represent the trend expected in objects of that kind. The solid black line is the bisector of the two lines and is used to partition points into one or other category, where they are colour-coded: blue for AGN and red for starburst-driven shocks.

The bottom part of Figure \ref{fig:bpt_dith} shows the colour-coded points re-mapped onto their spatial position in the galaxy. This helps to determine the spatial origin of the different ionisation mechanisms which are potentially in play. In all three IDD maps the disk of the galaxy is dominated, as expected, by star formation. However, the picture becomes more complex as we move off the plane of the disk. In all three IDDs an indication is given that a different, harder, ionisation source is at work away from the disk. The [\ion{N}{ii}] IDD and map are ambiguous in that they show evidence for both AGN excitation and shock excitation in the south-west and north-east parts of the galaxy.

The [\ion{S}{ii}] diagram is less ambiguous pointing towards both wind regions being dominated by ionisation from starburst driven shocks. The [\ion{O}{i}] diagram also hints more strongly towards shock ionisation over AGN ionisation. For these reasons the starburst-driven wind scenario is favoured, but it is not possible to entirely rule out a wind driven by an obscured AGN on the basis of the IDDs only.

The star formation rate (SFR) for ESO 185-G031 is derived from IRAS 60$\rm{\mu m}$ (F$_{60}$) and 100$\rm{\mu m}$ (F$_{100}$) flux measurements after \citet{Kewley02}. The measured values are F$_{60}$=0.52\,Jy and F$_{100}$=1.66\,Jy, yielding a value for far-infrared luminosity of L(FIR)=$1.92\times10^{43}$erg\,s$^{-1}$. This luminosity represents that over the range 40\,-\,120\,$\mu$m. The ratio of L(FIR) to L(IR), the latter being the total IR dust emission, is given by L(IR)=1.75\,L(FIR) \citep{Calzetti00}. This then gives a SFR of 1.7\,M$\rm{_{\odot}}$\,yr$^{-1}$ for ESO 185-G031.

The use of this conversion between L(FIR) and L(IR), and of L(IR) to calculate star formation rate, assumes that all dust emission is due to heating from the young stellar population \citep{Calzetti00}. If this is not the case, and some heating is from other sources such as an older population of stars, then this will overestimate the SFR. However, assuming the converse, that L(FIR) is a more accurate representation of the bolometric luminosity, will certainly underestimate the SFR. With this in mind we adopt 1.7\,M$\rm{_{\odot}}$\,yr$^{-1}$ as the correct SFR for this galaxy.

A second, independent, measurement of SFR can be made using ultraviolet emission. GALEX observations of this galaxy give $\rm{m_{FUV}}$=19.05~mag with E(B-V)=0.054~mag. Using the procedure in \citet{LS10} yields a value for SFR of 0.07\,M$\rm{_{\odot}}$\,yr$^{-1}$. A potential source of this discrepancy is the estimate of obscuration from the UV data. Given the brightness of this galaxy in FIR the true obscuration is likely to be larger, meaning the SFR estimate from the ultraviolet should be taken as a lower limit. The difference between these two indicators could also indicate the importance played by the star formation history of the galaxy in determining an accurate SFR \citep[e.g.][]{KLS09, LS+12}. However, individual estimates of SFR are still likely to be discrepant simply because the calibration of different SFR indicators can give estimates which vary by up to a factor of two (see for example \citet{Hopkins03}).

From the IMACS/MMTF imaging it is estimated that the star formation is occurring in the two prominent H$\alpha$ ``eyes'' covering an area roughly 3\,kpc in diameter. This implies a star formation surface density of about \,0.24\,M$_\odot$\,yr$^{-1}$\,kpc$^{-2}$. This is more than sufficient to launch a bipolar wind \citep{Heckman02}. In addition this galaxy does not appear in the SUMSS \citep{Mauch03} or AT20G \citep{Murphy10} catalogues implying that it is unlikely to host an AGN and the observed large-scale wind is therefore most likely driven by star formation.

An estimate of the metallicity of the star-forming disk was derived using only those spaxels showing a signature of star formation. We derived an integrated spectrum of the disk by masking the data cube based on the [\ion{N}{ii}]/H$\alpha$ line ratio, rejecting spaxels with [\ion{N}{ii}]/H$\alpha>-0.35$ and summing the resulting cube. We calculated values of [\ion{O}{iii}]/H$\beta$=0.577$\pm$0.048 and [\ion{N}{ii}]/H$\alpha$=0.349$\pm$0.018 and used the relations in \citep{Pettini04} to empirically determine the gas-phase oxygen abundance of this galaxy. We find a consistent value of 12+log(O/H)=8.64 and 8.66~dex using the N$_2$ and the O$_3$N$_2$ parameters, respectively.

It is important to note that these oxygen abundances correspond to values calibrated by following the direct method \citep{Osterbrock06}. To compare with oxygen abundances calculated using photoionization models, these values should be increased by 0.3-0.4~dex, i.e. the gas-phase metallicity of this galaxy is 12+log(O/H)$\sim$9.0 \citep[see][for a review of this problem]{LSD12}.

\subsection{Kinematics}

\subsubsection{The Pointed Data}

\begin{figure}[!h]
\centering
\includegraphics[width=8.0cm]{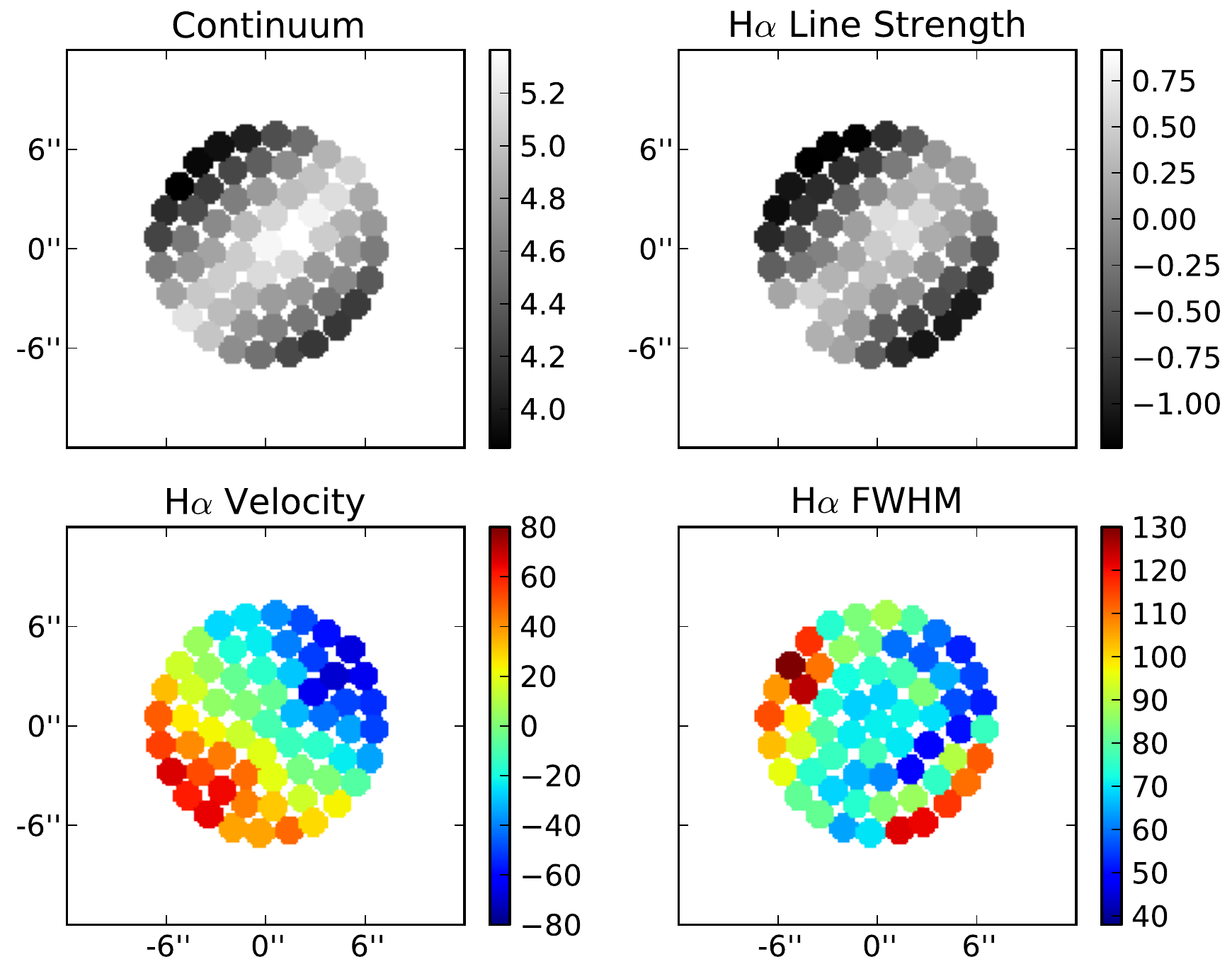}
\caption{The H$\alpha$ kinematics for ESO 185-G031 derived from the pointed observations. A kink is seen in the south-west of the velocity map and a trend towards broader line widths off the plane of the disk is seen in the FWHM map. These are both indicative of a complex kinematic structure comprised of more than one kinematically distinct component, supporting the presence of a wind in this galaxy.
\label{fig:kin_2hr}}
\end{figure}

Gas kinematics for the pointed data were derived using a Gaussian fit to the H$\alpha$ emission line in each spaxel. Figure \ref{fig:kin_2hr} shows the results, with a quasi-continuum map (summed from 6300\,-\,7300\,\AA) in the first panel, the H$\alpha$ line strength map in the second and the velocity and FWHM maps in the third and fourth respectively.

Close examination of the velocity field reveals some asymmetry, most notably a kink in the redshifted side of the velocity map, south of the plane of the disk.  The sigma map also shows a trend as one moves off the disk, with the emission line broadening in the gas off the plane of the disk. Broadening in this manner can be due to two distinct velocity components, such as a rotating disk and an outflow, superimposed along the line of sight, thus supporting the existence of a wind in this galaxy. 

\subsubsection{The Dithered Data}
\label{sec:kin_dither}
\begin{figure}[t]
\centering
\includegraphics[width=8.0cm]{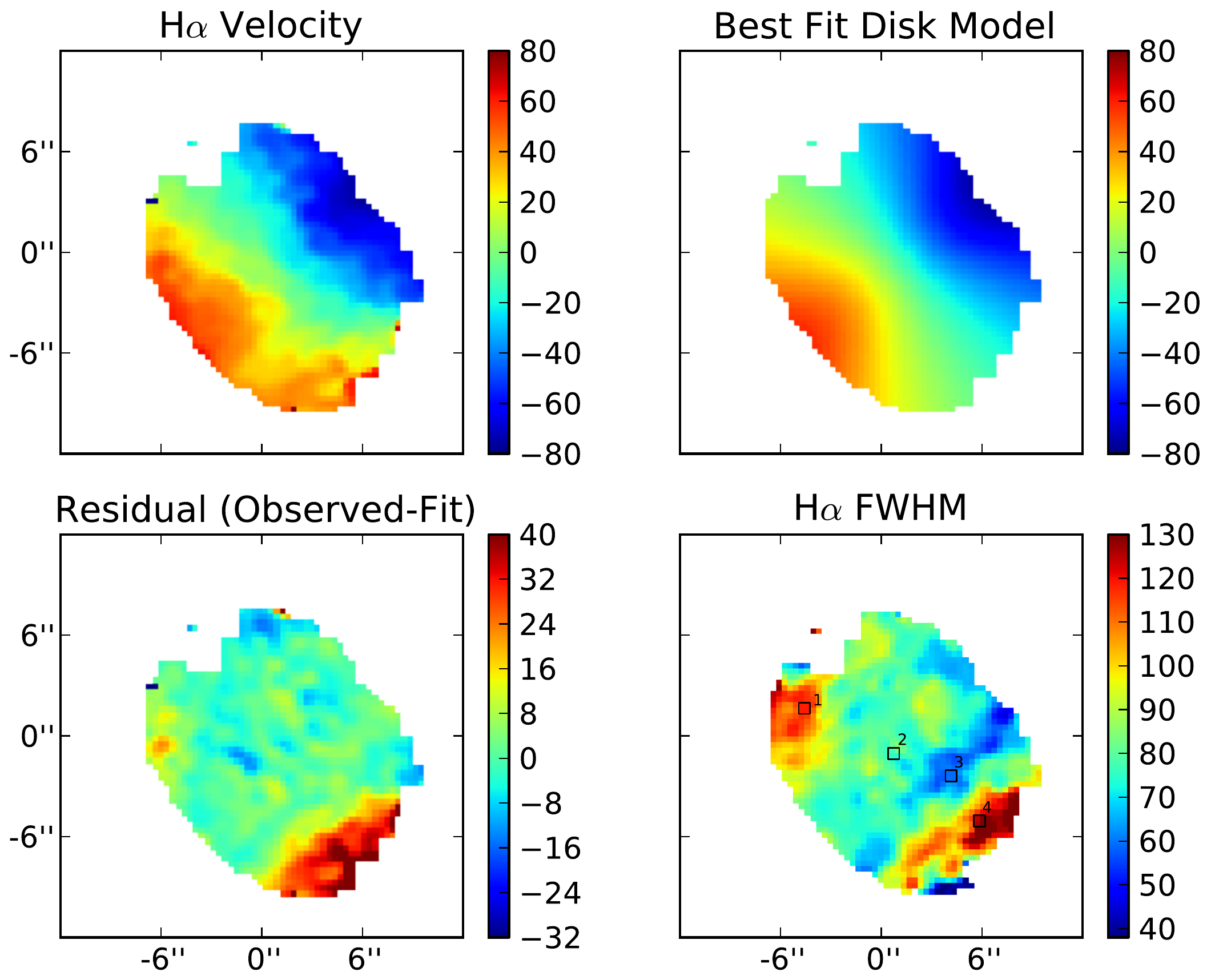}
\caption{The H$\alpha$ velocity map for ESO 185-G031 (top left) shown with the best fit disk model (top right; see Section \ref{sec:kin_dither} for details). The residual map, found by subtracting the model from the observed velocity map (bottom left), shows a very strong deviation from the rotating disk model in the region of the galactic wind. This region is co-spatial with the broadened emission lines seen in the FWHM map (bottom right). The squares overlaid on the FWHM map correspond to the four apertures for which individual H$\alpha$ spectra are displayed in Figure \ref{fig:spectra}.
\label{fig:kin_model}}
\end{figure}

For the dithered observations the H$\alpha$ kinematics are derived using the finely binned data cubes as the S/N in the H$\alpha$ emission line is high enough to fit in all spaxels. The dithered observations reveal the trends in the emission line kinematic properties much more convincingly. The top left panel of Figure \ref{fig:kin_model} shows the H$\alpha$ velocity field extracted from the dithered data. The structure in the map is very clear with the kink in the south of the disk readily apparent. The bottom right panel shows the derived FWHM of the H$\alpha$ emission line. The trend seen in the pointed data is reproduced, with significantly broader emission lines seen off the plane of the disk in both directions. The regions of the data with disturbed velocity and broadened emission lines correspond to those regions displaying non-star-formation driven ionisation properties.

\begin{figure}[t]
\centering
\includegraphics[width=8.0cm]{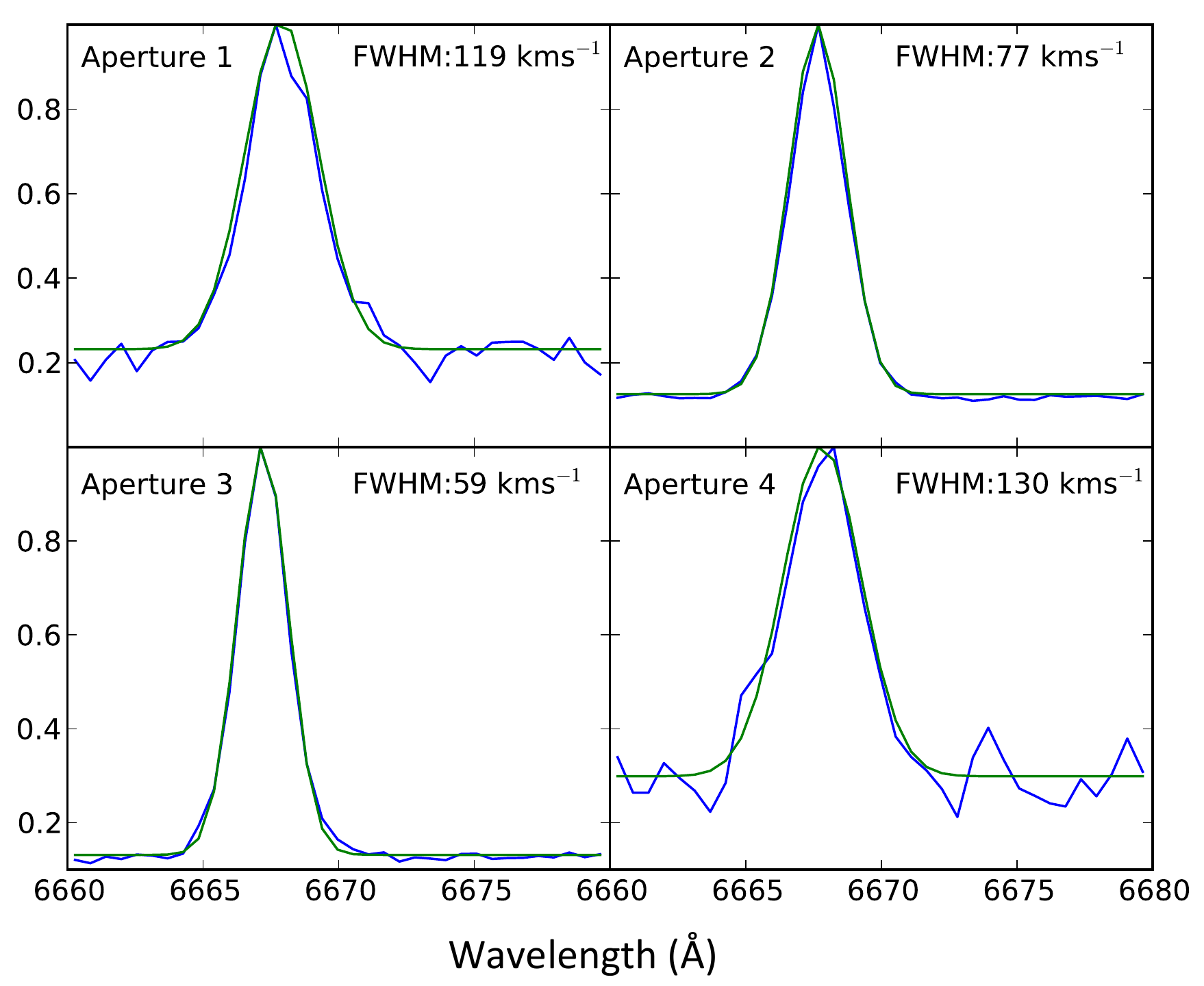}
\caption{Four spectra extracted from 2$\times$2 spaxel apertures. The extracted spectra are blue, with the single Gaussian fits shown in green. Apertures 2 and 3 in the disk show narrower Gaussian profiles, than apertures 1 and 4 (in the ``wind'' part of the galaxy) which show significantly broader profiles. This indicates that the emission in this region may come from two kinematically distinct components - the disk of the galaxy and the wind.
\label{fig:spectra}}
\end{figure}

The shape of the emission line changes off the plane of the disk as well, hinting at more than one velocity component in the gas (e.g. rotating disk and wind outflow). To illustrate this we extract spectra from four 2$\times$2 spaxel apertures, as shown in the bottom right panel of Figure \ref{fig:kin_model}. The corresponding spectra (zoomed in on the H$\alpha$ emission line) and Gaussian fits are shown in the four panels of Figure \ref{fig:spectra}. Apertures 2 and 3 are in the plane of the disk and show significantly narrower line profiles (FWHM=77$\pm7$ and 59$\pm3$\,km\,s$^{-1}$ respectively) than those seen in apertures 1 and 4 (FWHM=119$\pm13$ and 130$\pm16$\,km\,s$^{-1}$ respectively), which lie in the ``wind'' regions. This line-broadening supports the idea that we are observing two separate kinematic components superimposed, the rotating, star-forming disk and the wind outflow.

To further examine the nature of the observed velocity field a rotating disk model was used to fit the observations. Before the fit was performed the H$\alpha$ velocity map was masked using the [\ion{N}{ii}]/H$\alpha$ map to distinguish the parts of the map corresponding to star-forming disk. A threshold of [\ion{N}{ii}]/H$\alpha>-0.35$ effectively masks the parts of the galaxy showing non-star-forming ionisation.

We use a simple disk model to fit the projected H$\alpha$ velocities. Our model uses an arctangent rotation curve \citep{Staveley90}, given by 
\begin{equation}
  V(r) = \frac{2 V_{\rm asym}}{\pi} \arctan \left( \frac{r}{r_d} \right)
\end{equation}
where $V_{\rm asym}$ is the asymptotic circular velocity, and $r_d$ is the kinematic scale radius.  The position angle and inclination determine the 3D spatial orientation of the disk. The model also includes the position of the centre of the disk within the velocity map, and a systematic velocity offset (which corrects for any bias in the systemic redshift) as free parameters. The model produces a velocity map analogous to the observed map. A Levenberg-Marquardt minimisation routine optimises the model's seven free parameters to give the best fitting model velocity map.

The results are shown in Figure \ref{fig:kin_model}, with the disk model displayed in the top right panel and the difference between the observed velocity map and the model in the bottom left panel. The residual map shows the region in the south-west of the observations contains gas not rotating with the disk. We interpret this as material flowing away from us in an outflow. The outflow may in fact be bipolar as is hinted in the H$\alpha$ FWHM map but we do not have enough spatial coverage to the north of the galaxy to confirm this. Of note as well is the sharpness of the velocity transition between the rotating disk and the extra-planar material. This is thought to be due to the inclination of the disk, which is derived from the velocity model to be 63 degrees. With the disk inclined thus to the line of sight the dusty edge of the disk is seen to the south-west of the disk (see Figure \ref{fig:im_mmtf}) and probably corresponds to the region of low FWHM in the H$\alpha$ kinematics (see panel 4 of Figure \ref{fig:kin_model}). 

\begin{figure}[t]
\centering
\includegraphics[width=8.0cm]{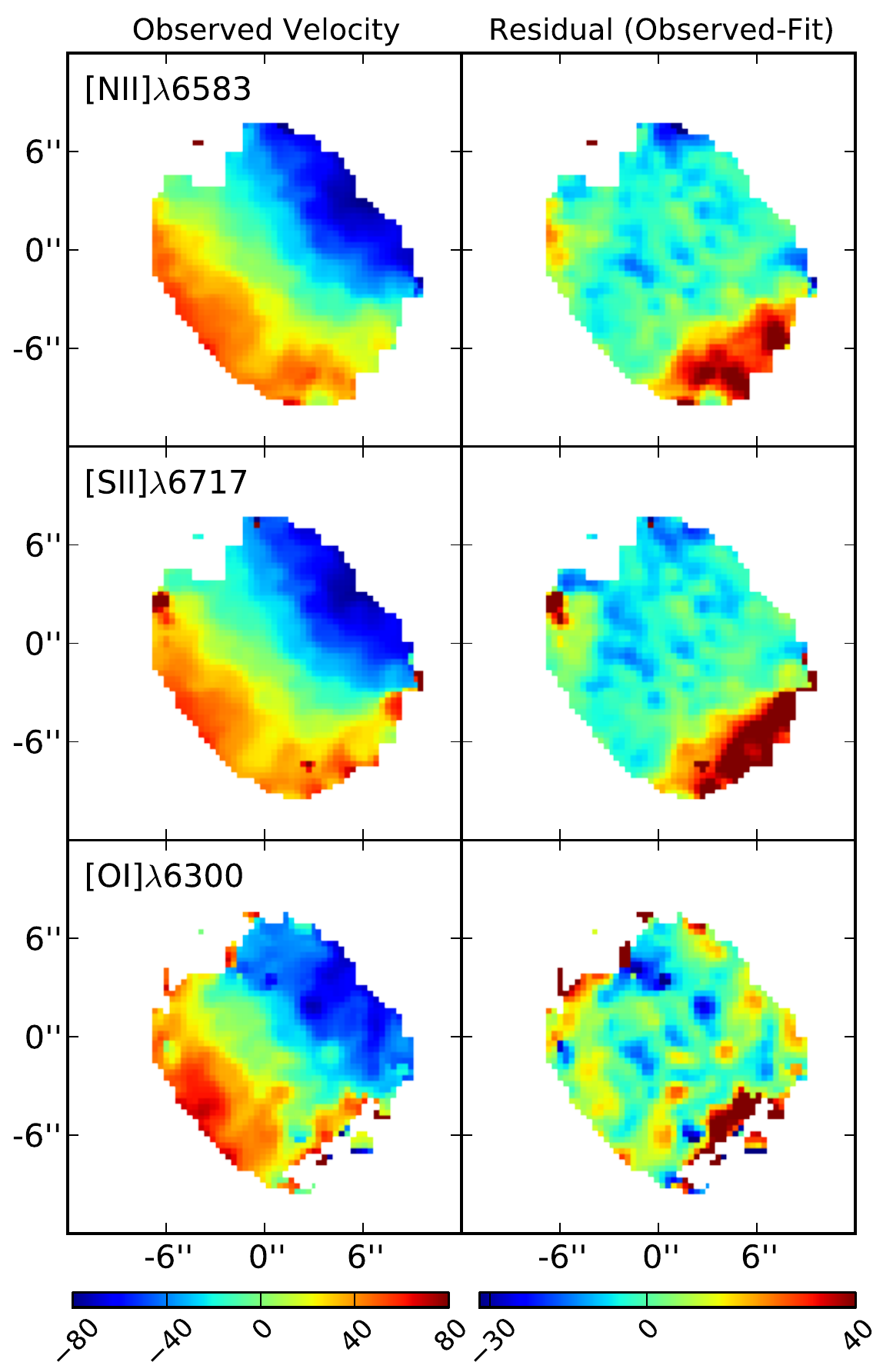}
\caption{The left-hand column shows the velocity maps derived for each of the [\ion{N}{ii}]$\lambda$\,6583, [\ion{S}{ii}]$\lambda$\,6717 and [\ion{O}{i}]$\lambda$\,6300\,\AA\, emission lines. The right-hand column shows the residuals found when the disk fit to the H$\alpha$ velocity map is subtracted from each of the maps. The wind signature is clearly seen in each of the residual maps.
\label{fig:red_vel_res}}
\end{figure}

\begin{figure}[t]
\centering
\includegraphics[width=8.0cm]{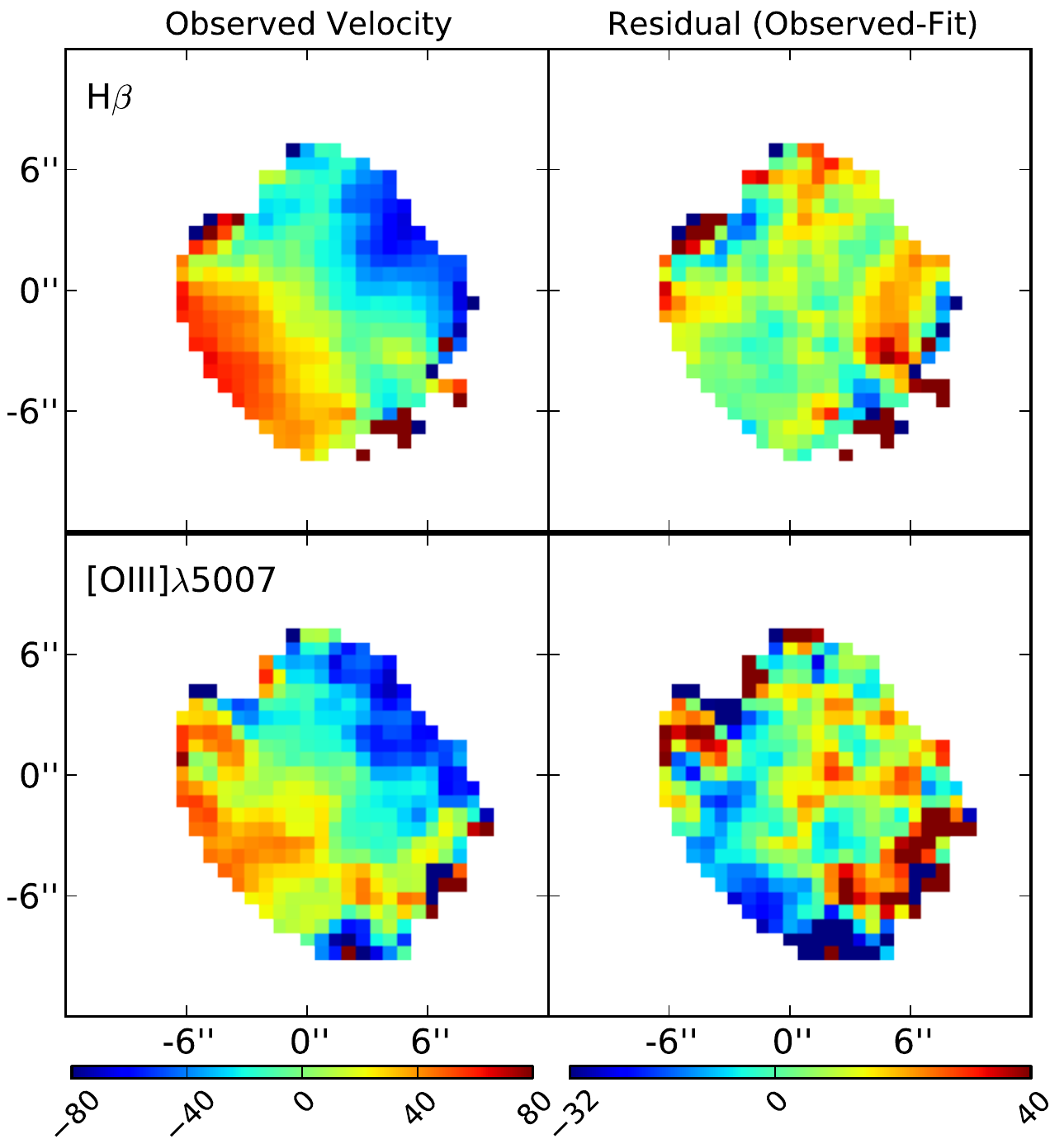}
\caption{The left-hand column shows the velocity maps derived for the H$\beta$ and [\ion{O}{iii}]$\lambda$\,5007\,\AA\, emission lines. The right-hand column shows the residuals found when the disk fit to the H$\alpha$ velocity map is subtracted from each of the maps. The wind signature is clearly seen in each of the residual maps.
\label{fig:blue_vel_res}}
\end{figure}

The fit to the H$\alpha$ kinematics is taken to represent the genuine rotation of the gas in the galaxy. Velocity maps are derived from all other emission lines and the disk model is subtracted from each in turn to examine deviations from the rotating disk. For the red emission lines, [\ion{O}{i}]$\lambda$\,6300, [\ion{N}{ii}]$\lambda$\,6583 and [\ion{S}{ii}]$\lambda$\,6717\,\AA\, the finely binned cubes are used. The results are shown in Figure \ref{fig:red_vel_res}. In all three velocity maps the signature of the wind is clear. For the blue emission lines the more coarsely binned cubes are use to derive the kinematic properties (with the disk model binned accordingly before subtracting). The results are shown in Figure \ref{fig:blue_vel_res}. Again, the wind signature is seen in the south-west of the residual maps.

\subsection{Imaging}
\label{sec:im_analysis}

\begin{figure}[!ht]
\centering
\includegraphics[width=8.0cm]{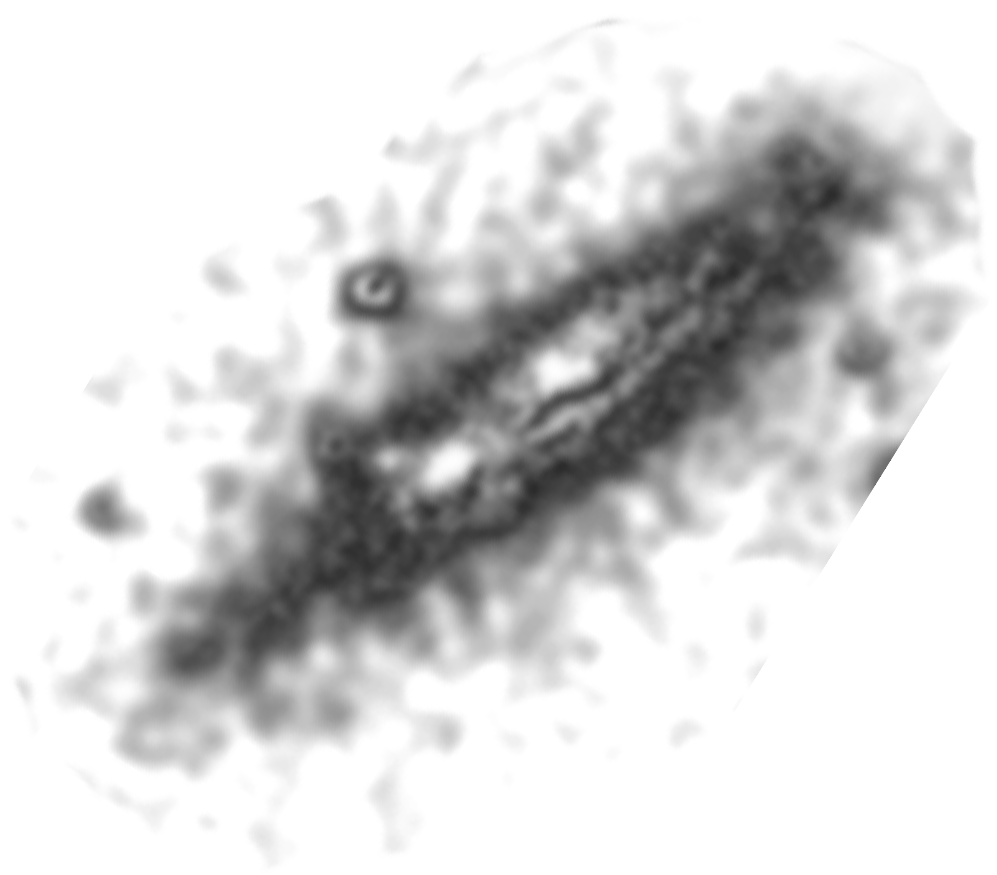}
\caption{A $31\arcsec\times 54\arcsec$ region extracted from the MMTF image, where north points upwards and east to left. This continuum subtracted emission-line-only image is a composite, with inverted gray scale to show the outer parts of the galaxy smoothed to 0\farcs{8} FWHM and positive gray scale showing the disk at 0\farcs{6}. Numerous faint H$\alpha$+[\ion{N}{ii}] filaments extend up to 11$\arcsec$ (3.7 kpc) from the disk.
\label{fig:im_mmtf_deep}}
\end{figure}

For the IMACS/MMTF image analysis, each of the individual frames were aligned on stars, scaled to a constant sky value, and the stack mean was used to form the science image except where the brightest pixel in the stack deviated strongly from the rest so was replaced by the median before taking the stack mean. Seven cycles of Lucy-Richardson deconvolution were applied to the line image to match its stellar PSF to that of the continuum image. The continuum was scaled to produce zero subtraction residuals within the galaxy; a scale change of 2.7\% between line and continuum was also applied. Foreground stars are still evident because of the MMTF passband used here. In Fig.~\ref{fig:im_mmtf_deep},  two H$\alpha$-bright ``eyes'' straddle the galaxy center in the disk, while many faint filaments extend above the disk. This extended emission strongly supports the discovery of a wind in this galaxy.

\section{Conclusions}
\label{sec:conc}

We have presented SAMI IFU observations of a single galaxy observed in July 2011. This galaxy, ESO 185-G031, was part of the commissioning sample of 13 and showed interesting structure in its kinematic and optical line ratio maps. All four of the measurable line ratios show a significant trend to higher values off the plane of the disk. Using IDDs to further investigate this reveals that emission in the disk is dominated by ionisation from star formation, but the regions off the plane of the disk exhibit ionisation by some other mechanism. We propose that this other mechanism is shock excitation from a starburst-driven wind, based on comparison with the IDDs for NGC1482, a canonical starburst driven wind \citep{SBH10}.

The galaxy also shows deviations from a simple rotating disk in its spatially resolved velocity field, implying that more than one kinematically distinct processes are at play. This is borne out when comparing the measured velocity field to a disk model fit to the part of the velocity field corresponding to the star forming disk (as defined by the [\ion{N}{ii}]/H$\alpha$ emission line ratio). A clear velocity residual is seen which is co-spatial with the region of elevated line ratios, supporting to the wind hypothesis for this galaxy.

Previous studies have identified winds through gas kinematics using long-slit spectroscopy \citep{Axon78} or through variations in gas excitation using tunable filters \citep{Veilleux02}. Here we demonstrate that integral field spectroscopy is able to identify galactic winds through a combination of gas excitation and kinematics. Thus we anticipate that a large survey using hexabundles will enable the first objective survey of the prevalence of outflows in galaxies.

More generally, SAMI will be very effective at demonstrating departures from symmetry over a wide range of galaxy types, both in the stars and in the ionised gas. We anticipate that all or most galaxies will reveal such behaviour, as evidenced by the recent SAURON \citep{deZeeuw02} and ATLAS-3D \citep{ATLAS-3D} surveys. Future samples will be so large that we can study this behaviour as a function of  environment across the hierarchy \citep{Lawrence12}. Just how galaxies are shaped by their environments is one of the great unknowns of modern astronomy. We fully anticipate that hexabundle survey spectrographs will have a major impact in answering this question in the years ahead.

\begin{acknowledgements}
We thank the staff of the Australian Astronomical Observatory and those at the University of Sydney for their excellent support in developing, commissioning and operation of the  SAMI instrument.

Parts of this research were conducted by the Australian Research Council Centre of Excellence for All-sky Astrophysics (CAASTRO), through project number CE110001020.  SMC acknowledges the support of an Australian Research Council Future Fellowship (FT100100457). JBH is supported by a Federation Fellowship from the ARC. CQT gratefully acknowledges support by the National Science Foundation Graduate Research Fellowship under Grant No. DGE-1035963.
\end{acknowledgements}

\end{document}